\documentclass[%
reprint,
superscriptaddress,
nobibnotes,
amsmath,amssymb,
aps,
floatfix
]{revtex4-2}

\usepackage[T1]{fontenc} 
\usepackage[]{graphicx}
\usepackage[utf8]{inputenc}
\usepackage{blindtext}
\usepackage{lmodern}%
\usepackage{hyperref}%
\usepackage{xcolor}%
\usepackage{placeins}
\usepackage{bm}
\usepackage{tabularx}
\usepackage{booktabs}
\usepackage{notes2bib}
\usepackage[normalem]{ulem}
\usepackage{SIunits}

\newcommand{\VIThree}{VI\textsubscript{3}}
\newcommand{\CrIThree}{CrI\textsubscript{3}}
\newcommand{\CrClThree}{CrCl\textsubscript{3}}
\newcommand{\CrBrThree}{CrBr\textsubscript{3}}
\newcommand{\Tc}{\ensuremath{T_{\rm C}}}
\newcommand{\muH}{\ensuremath{\mu_0 H}}
\newcommand{\muHperp}{\ensuremath{\mu_0 H_\perp}}
\newcommand{\muHparallel}{\ensuremath{\mu_0 H_\parallel}}

\newcommand{\SiOTwo}{SiO\textsubscript{2}}
\newcommand{\Isd}{\ensuremath{I_{\rm {sd}}}}
\newcommand{\Vsd}{\ensuremath{V_{\rm {sd}}}}
\newcommand{\VG}{\ensuremath{V_{\rm {g}}}}

\newcommand{\mum}{$\rm{\mu}$m}

\renewcommand{\degree}{\ensuremath{^\circ\,}}

\newcommand{\dqmp}{Department of Quantum Matter Physics, University of Geneva, 24 Quai Ernest Ansermet, CH-1211 Geneva, Switzerland}
\newcommand{\gap}{Department of Applied Physics, University of Geneva, 24 Quai Ernest Ansermet, CH-1211 Geneva, Switzerland}

\newcommand{\Cambridge}{Department of Physics, Massachusetts Institute of Technology, Cambridge, Massachusetts 02139, USA}

\newcommand{\ie}{\emph{i.e.},}

\definecolor{AAcolor}{rgb}{0.7,0.1,0.4}

\usepackage{newfloat}

\definecolor{linkcol}{rgb}{0,0,0.4}
\definecolor{citecol}{rgb}{0.5,0,0}
\definecolor{harvardcrimson}{rgb}{0.79, 0.0, 0.09}
\definecolor{lava}{rgb}{0.81, 0.06, 0.13}

\hypersetup{
	bookmarksopen=true,
	pdftitle="Probing magnetism in exfoliated VI3 layers with magnetotransport",
	pdfauthor="Soler-Delgado et al",
	pdfsubject="Probing magnetism in exfoliated VI3 layers with magnetotransport measurements", 
	pdfstartview={FitH}, 
	pdfmenubar=true, 
	pdfhighlight=/O, 
	colorlinks=true, 
	pdfpagemode=UseNone, 
	pdfpagelayout=SinglePage, 
	pdffitwindow=true, 
	linkcolor=harvardcrimson, 
	citecolor=harvardcrimson, 
	urlcolor=lava
}

\begin{document}

	\author{David Soler-Delgado}\affiliation{\dqmp} \affiliation{\gap} 
	
	\author{Feng-rui Yao}\affiliation{\dqmp} \affiliation{\gap}     

    \author{Dumitru Dumcenco} \affiliation{\dqmp}       

	\author{Enrico Giannini} \affiliation{\dqmp}        
	
 	\author{Jiaruo Li } \affiliation{\Cambridge}        
 	
 	\author{Connor A. Occhialini} \affiliation{\Cambridge}  
 	\author{Riccardo Comin} \affiliation{\Cambridge}        

	
	
	\author{Nicolas Ubrig} \affiliation{\dqmp} \affiliation{\gap} 
	
	\author{Alberto F. Morpurgo} \email{alberto.morpurgo@unige.ch}\affiliation{\dqmp} \affiliation{\gap}

	
	\date{\today}
	
	
	\title{Probing magnetism in exfoliated \VIThree\ layers with magnetotransport}
	
	\begin{abstract} 
    We perform magnetotransport experiments on \VIThree multilayers, to investigate the relation between ferromagnetism in bulk and in exfoliated layers. The magnetoconductance measured on field-effect transistors and tunnel barriers shows that the Curie temperature of exfoliated multilayers is \Tc~=~57~K, larger than in bulk ($T_{\rm C,bulk}$~=~50~K). Below $T~\approx$~40~K, we observe an unusual evolution of the tunneling magnetoconductance, analogous to the phenomenology observed in bulk. Comparing the magnetoconductance measured for fields applied in- or out-of-plane corroborates the analogy, allows us to determine that the orientation of the easy-axis in multilayers is similar to that in bulk, and suggests that the in-plane component of the magnetization points in different directions in different layers. Besides establishing that the magnetic state of bulk and multilayers are similar, our experiments illustrate the complementarity of magnetotransport and magneto-optical measurements to probe magnetism in 2D materials.
    \end{abstract}
	\maketitle

Atomically thin layers of many different materials have been produced by exfoliating bulk crystals of van der Waals bonded compounds~\cite{novoselov_two-dimensional_2005,mounet_two-dimensional_2018}. The crystalline structure of exfoliated layers is commonly assumed to be the same as that of the bulk parent crystals, and indeed experiments normally confirm this assumption. This is however not the case for many recently discovered atomically thin magnetic materials~\cite{gong_discovery_2017,huang_layer-dependent_2017,burch_magnetism_2018,gong_two-dimensional_2019,gibertini_magnetic_2019,mak_probing_2019,huang_emergent_2020}, whose structure in  exfoliated form  differs from that of the bulk, often resulting  in drastically different magnetic properties~\cite{sun_giant_2019,klein_enhancement_2019,ubrig_low-temperature_2019,mccreary_distinct_2020,gibertini_magnetism_2020}. Examples are  provided by CrI$_3$ --whose multilayers  are layered antiferromagnet with \Tc~=~51~K whereas bulk crystals are ferromagnets with \Tc~=~60~K~\cite{dillon_magneto-optical_1966,huang_layer-dependent_2017,wang_very_2018}-- and  CrCl$_3$ --in which the interlayer exchange interaction in multilayers is approximately one order of magnitude larger than in the bulk~\cite{wang_determining_2019}.\\

\begin{figure}[h!]
    \centering
    \includegraphics{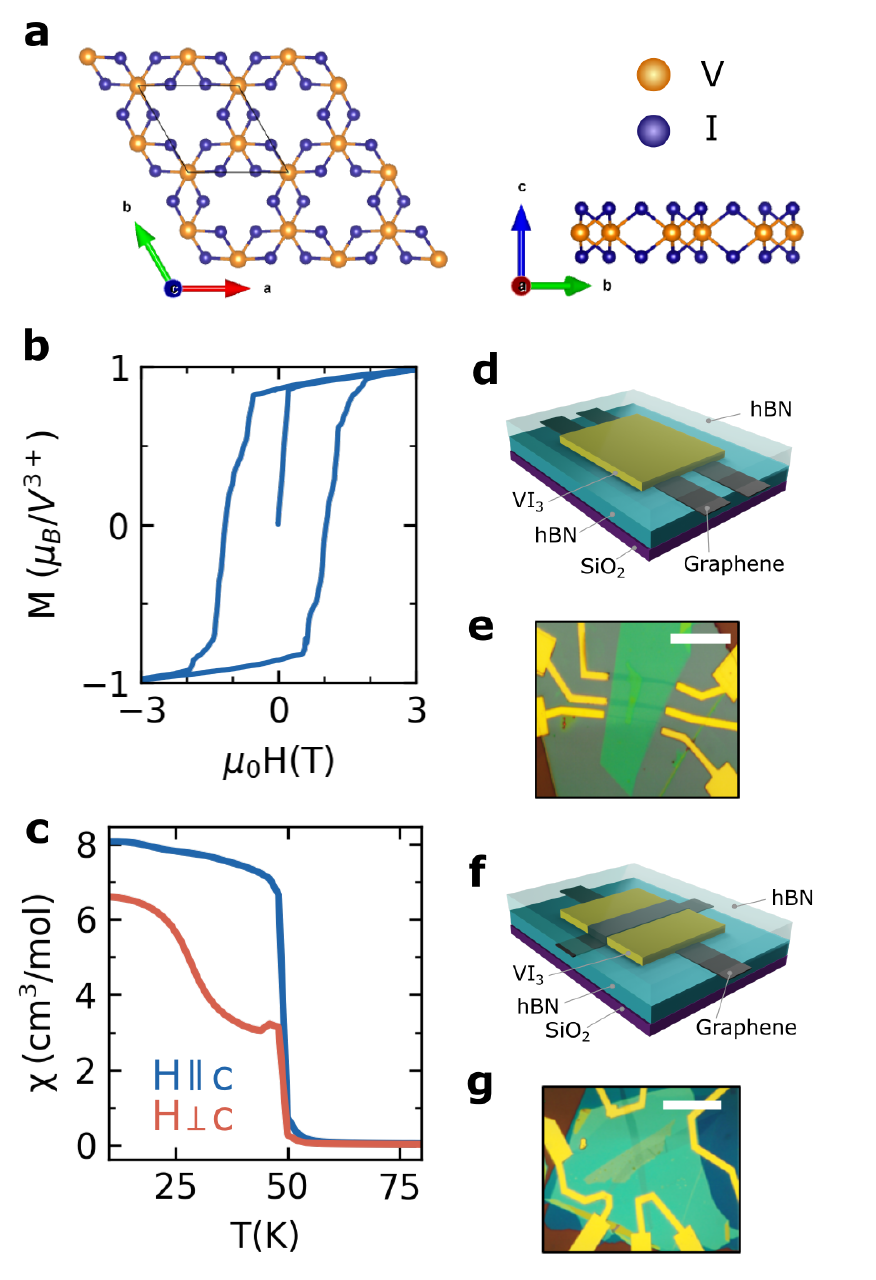}
    \caption{\textbf{(a)} Top (left) and side (right) view of the crystal structure of a  \VIThree\ monolayer. The purple and orange balls represent the iodine and vanadium atoms, respectively. \textbf{(b)} Magnetization, $M(H)$, of  \VIThree\ bulk crystals measured at $T~=~2$~K as function of magnetic field applied perpendicularly to the layers, showing hysteretic behavior. \textbf{(c)} Field-cooled susceptibility $\chi(T)$ (\muHperp~=~10~mT) of our bulk \VIThree\ crystals, measured in a temperature range between 5~K~<~$T$~<~80~K. The blue and red curves denote measurement configurations where the magnetic field is applied parallel to the c-plane and ab-plane, respectively.  Schematic representation \textbf{(d)} of a \VIThree\ transistor with graphene contacts and optical micrograph \textbf{(e)} of an actual device based on a 18~nm thick multilayer. The highly doped Silicon substrate covered by a 285~nm \SiOTwo\ layer is used as gate (the scale bar is 10~\mum\ long). Schematic representation \textbf{(f)} and optical micrograph \textbf{(g)} of a tunneling device, with graphene contacts on opposite sides of a 18~nm thick \VIThree\ multilayer (the scale bar is 10~\mum\ long). All structures are encapsulated between hexagonal boron nitride (hBN) crystals to avoid degradation of \VIThree.}
    \label{fig:1}
\end{figure}

\VIThree\ (see Fig.~\ref{fig:1}a) is an example of current interest, which exhibits conspicuous structural differences in bulk and exfoliated multilayers, accompanied by a  magnetic response that  appears to be strikingly different in the two cases. Bulk \VIThree\  crystals possess inversion symmetry and at \Tc~=~50~K undergo a transition  into a ferromagnetic state, suggested to be of the Ising type~\cite{son_bulk_2019,lin_magnetism_2021}. Recent experiments, however, showcase an unusual evolution of the magnetic properties  below  40~K, which is indicative of a  more complex magnetic state~\cite{ dolezal_crystal_2019,koriki_magnetic_2021,gati_multiple_2019, son_bulk_2019,kong_vi_2019,tian_ferromagnetic_2019,valenta_pressure-induced_2021,marchandier_crystallographic_2021}. Experimental observations include the splitting of diffraction peaks in neutron and X-Ray spectroscopy~\cite{marchandier_crystallographic_2021,dolezal_crystal_2019}; the onset of a disproportionation between the (supposedly) structurally equivalent V atoms, detected by nuclear magnetic resonance~\cite{gati_multiple_2019}; a pronounced increase in the in-plane magnetic susceptibility found in magnetization measurements \cite{son_bulk_2019,tian_ferromagnetic_2019, valenta_pressure-induced_2021} (see Fig.~\ref{fig:1}c); a 30 degree rotation around the direction normal to the layers of the six-fold symmetric in-plane easy axis, which occurs upon cooling between 40 and 30~K~\cite{koriki_magnetic_2021}. No magnetic state capable to explain all the observed phenomenology has been proposed.\\

The fewer experiments \cite{lyu_probing_2020,broadway_imaging_2020,lin_magnetism_2021} reported on exfoliated layers indicate that the  crystalline structure lacks inversion symmetry (for trilayers or thicker layers), and that  a ferromagnetic state occurs below \Tc~=~57~K, a value  significantly higher than  the bulk \Tc~, in contrast to expectations~\cite{huang_layer-dependent_2017,gong_discovery_2017,kim_evolution_2019}. Both scanning magnetometry and reflective magnetic circular-dichroism (RMCD) measurements~\cite{broadway_imaging_2020,lin_magnetism_2021} exhibit a behavior consistent with Ising ferromagnetism at all temperatures, with no anomalies. It, therefore, appears that the behavior of multilayers differs --and is simpler-- than that observed in the bulk, but it is unclear whether this conclusion is just a consequence of the few experimental techniques available that offer enough sensitivity to probe magnetism in thin layers.\\

Here, we investigate exfoliated \VIThree\ layers by means of magnetotransport experiments and show that  --despite their different structural properties and larger critical temperature-- their magnetic response is similar to the one observed in bulk crystals, and not consistent with that of an  Ising ferromagnet \cite{wang_magnetization_2021}. Magnetotransport was measured using field-effect devices (Fig.~\ref{fig:1}d-e) and in devices in which  \VIThree\ multilayers act as tunnel barriers (Fig.~\ref{fig:1}f-g). In both configurations, a ferromagnetic transition is observed at \Tc~=~57~K, consistently with earlier RMCD measurements~\cite{lin_magnetism_2021}. Nevertheless,  the tunneling magnetoconductance below \Tc\ deviates qualitatively from that of an Ising ferromagnet and exhibits a behavior in line with that of bulk crystals. Specifically,  below 40~K, the magnetoconductance measured in a magnetic field perpendicular to the planes becomes negative  (whereas it is always positive in an established Ising ferromagnet such as \CrBrThree\ ~\cite{wang_magnetization_2021}), and the magnetoconductance in an in-plane field becomes pronouncedly hysteretic. By analyzing the magnetoconductance measured with the field applied in the two directions, we determine that the easy axis in \VIThree\ multilayers forms an angle of approximately 30\degree\ with respect to the normal to the \VIThree\ planes. Such angle is close to --but smaller than-- the that observed in bulk crystals. Besides showing that the magnetic state of exfoliated  \VIThree\ multilayers and bulk crystals are similar, our results establish that the anomalous magnetic response originates from the in-plane component of the magnetization and  illustrate the complementarity of magneto-optical and magneto-transport measurements to probe 2D magnetic materials.\\

The fabrication of \VIThree\ devices relies on micromechanical exfoliation of bulk crystals (characterized by magnetization and susceptibility measurements; see Fig.~\ref{fig:1}b and \ref{fig:1}c) to obtain multilayers. The multilayers are processed to form field-effect transistors and tunnel barriers (see Fig.~\ref{fig:1}d-g), using conventional pick-up and transfer techniques based on polymeric stamps~\cite{zomer_transfer_2011}. In practice, \VIThree\ multilayers are contacted with multilayer graphene strips and encapsulated in between exfoliated hBN layers ($\approx$~20-50~nm thick) to avoid degradation (exfoliation and assembly of the structures are carried out in the controlled atmosphere of a glove box). We attach metal contacts to the graphene strips using conventional electron-beam lithography in combination with reactive ion etching, evaporation of a Cr/Au film, and lift-off. All structures are realized on highly doped Si substrates (acting as gates in transistor devices) covered with 285~nm \SiOTwo.  We have investigated two transistors and four tunnel junction devices that exhibit fully consistent behavior, and here we present representative data from a selected transistor and a selected tunnel junction (See the Supporting~Information~section~S2 for data of additional devices).\\

\begin{figure*}[h!]
    \centering
    \includegraphics[width=\textwidth]{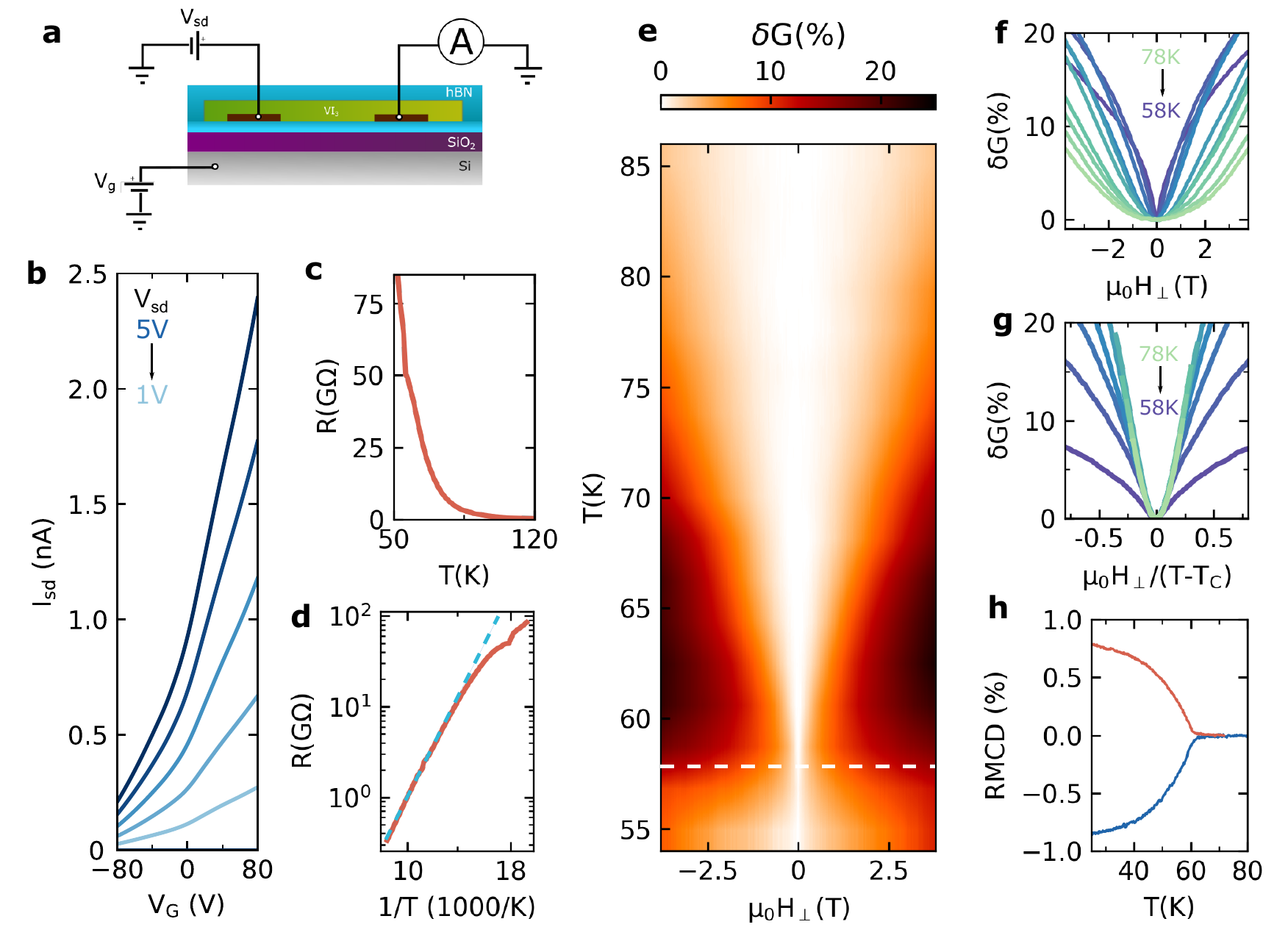}
    \caption{ \textbf{(a)} Schematic illustration of a field-effect transistor device used in the experiments. \textbf{(b)} Transfer curve (\Isd\-vs-\VG) as the source-drain voltage is \Vsd\ is varied from 5~V to 1~V in 1~V step.  \textbf{(c)} Temperature dependence of the resistance, $R(T)$, of a \VIThree\ transistor measured with \Vsd~=~+8~V and \VG~=~+100~V, exhibiting a pronounced increase upon cooling  \textbf{(d)} logarithmic scale plot of $R(T)$ as a function of $1/T$. The linear dependence implies a thermally activated behavior with activation energy of $E_a~\approx$~50~meV, as determined from a linear  fit (blue dashed line). \textbf{(e)} False color plot of the in-plane magnetoconductance $\delta G (H_{\perp},T) = (G(H_{\perp},T)-G(0,T))/G(0,T)$ as a function of the magnetic field, \muH$_{\perp}$, applied along the $c$-axis and temperature $T$ (the plot shows an interpolation of data measured at every 2~K as a function of $\mu_0H_{\perp}$). The white dashed line indicates the ferromagnetic transition at \Tc~=~57~K.  \textbf{(f)} Magnetoconductance $\delta G$ for $T$~>~\Tc, as $T$ is varied from 78~K to 58~K, in 2~K steps. \textbf{(g)} Same data shown in (f) plotted as a function of $\muH_{\perp}/(T-\Tc)$. When plotted in this way, all curves collapse on top of each other at small $H_{\perp}$. \textbf{(h)} Temperature dependence of the RMCD signal measured on a \VIThree\ tunnel barrier device based on a 10~nm thick \VIThree\ layer. The RMCD was measured in the region between the graphene tunnelling electrodes, while warming the sample up in the absence of a magnetic field, after having cooled down the device down to $\sim$~20~K in an applied perpendicular magnetic field of either  +980~mT (red line) or -980~mT (blue line).}
    \label{fig:2}
\end{figure*}

We first discuss transport measurements performed on a \VIThree\ transistor (See  Fig.~\ref{fig:2}a). Fig.~\ref{fig:2}b shows the transfer curves (\ie\ source-drain current \Isd\ as a function of gate voltage \VG), measured at $T$~=~90~K, as \Vsd\ is varied from 5~V to 1~V. The application of a positive gate voltage --corresponding to accumulating electrons at the surface of \VIThree-- causes a large increase in current. Even at the largest positive gate voltage, however, the low-temperature resistance is extremely high and increases in a thermally activated fashion upon cooling (see Fig.~\ref{fig:2}c-d), indicating that the accumulated electrons are localized and that transport is mediated by hopping (the activation energy --approximately 50~meV at \VG~=~+100~V-- corresponds to the distance in the energy of the localized electrons at the Fermi level and the conduction band). Under these conditions, it is unclear whether any magnetoconductance can be measured. \\

A magnetoconductance $\delta G(H_\perp,T) = (G(H_\perp,T)-G(0,T))/G(0,T)$  is nevertheless present, and exhibits a systematic evolution as a function of  magnetic field applied perpendicular to the layers, \muHperp, and temperature $T$ (Fig.~\ref{fig:2}e). The magnetoconductance sets in at a lower magnetic field as the temperature $T$ is lowered from 75~K to 58~K (see the horizontal dashed line),  a manifestation of the critical regime in the paramagnetic state of \VIThree\ near the ferromagnetic transition. Specifically, for a ferromagnet the conductance is expected to increase as the \VIThree\ magnetization increases, \ie\ as the spins in the material orient themselves in the same direction. Since the magnetic susceptibility $\chi \propto 1/(T-\Tc)$ diverges at \Tc, a smaller value of $\mu_0 H_\perp$ is needed to generate the same magnetization as $T$ approaches \Tc, explaining why the magnetoconductance sets in at lower fields. Indeed, if we plot the magnetoconductance measured at different temperatures (Fig.~\ref{fig:2}f) as a function of $\muHperp /(T - \Tc)$ (which in the linear regime is proportional to the magnetization) all curves overlap at small $H_\perp$ (Fig.~\ref{fig:2}g). We conclude that --in the paramagnetic state-- the magnetoconductance is a function of the magnetization and that the susceptibility tends to diverge as $T$ approaches \Tc. Optimizing the scaling of the magnetoconductance curves allows us to determine \Tc\ to be 57~K (or more precisely between 56~K and 58~K; the precision of the method originates from the 2~K temperature interval with which data are taken). \\

\begin{figure}[h!]
    \centering
    \includegraphics[width=0.49\textwidth]{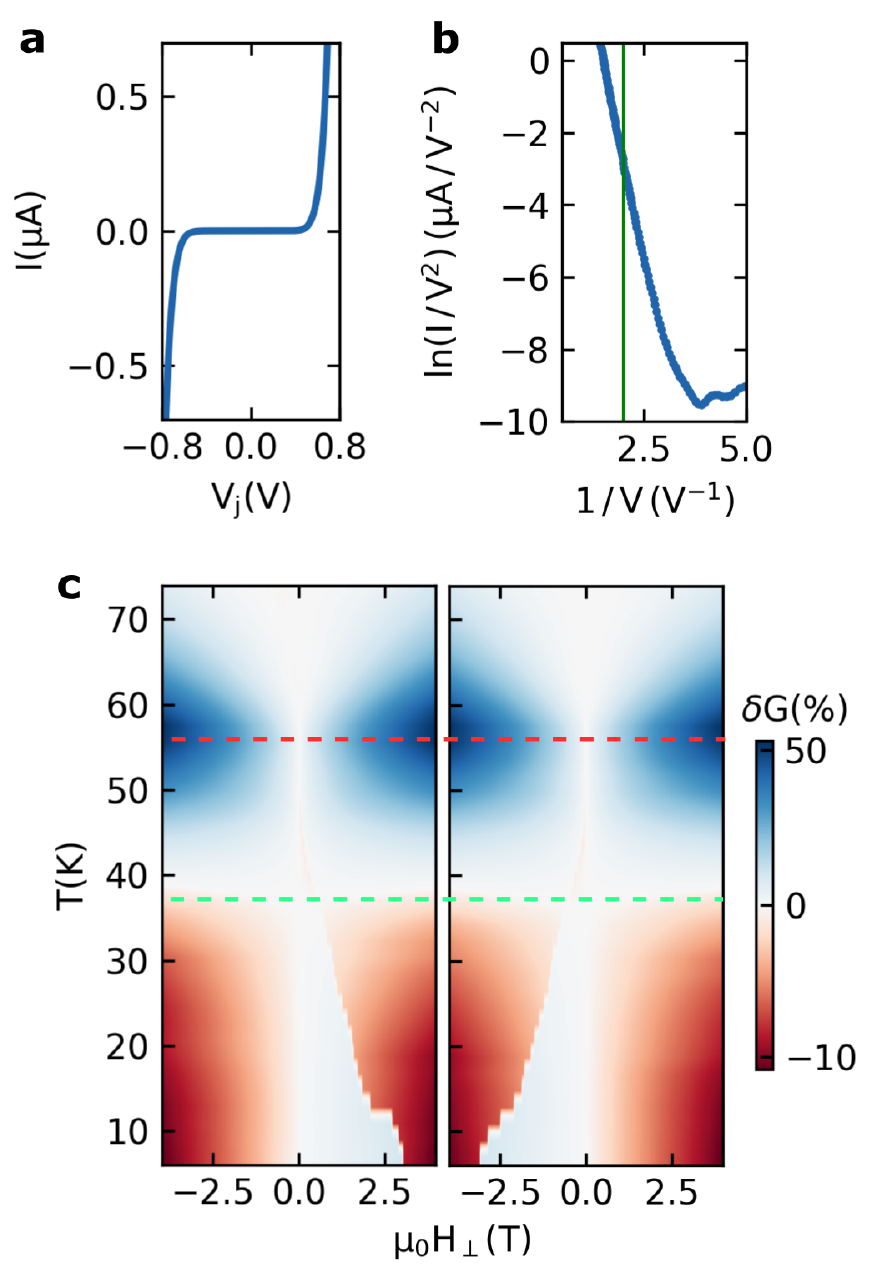}
    \caption{ \textbf{(a)} Current, $I$, across the tunnel barrier measured at $T$~=~2~K as a function of voltage. \textbf{(b)} Similar to what is found in barriers made of different Chromium trihalides, at sufficiently large bias the current behaves as expected for  Fowler-Nordheim tunneling, with  $\ln(I/ V^2)$  linearly proportional to $1/V$.  The green solid line ($V~=~0.5$~V) indicates the voltage at which the magnetoconductance data show in  \textbf{(c)} (and in Fig.~\ref{fig:4}) were measured. \textbf{(c)} Color plot of tunneling magnetoconductance as a function of applied field \muHperp\ and temperature $T$. Left and right panels represent the magneto-conductance measured while sweeping the field from -4~T to +4~T, or from +4~T to -4~T, respectively. The red dashed line in both panels indicates the Curie temperature \Tc\ determined from the magnetoconductance data. The green dashed line in both panels denotes the temperature at which the magnetoconductance becomes negative ($T~\approx~36$~K).}
    \label{fig:3}
    \end{figure}

In-plane transport measurements, therefore, confirm that exfoliated multilayers have a significantly higher \Tc\ ~than the bulk. This value of the Curie temperature agrees well with the one determined in recent RMCD experiments~\cite{lin_magnetism_2021}, as well as in our own. Fig.~\ref{fig:2}h shows the temperature dependence of the circular-dichroism of the light reflected from a \VIThree\ exfoliated layer, measured at zero applied magnetic field, after having cooled down the multilayer to 20~K in either a +980~mT (red curve) or -980~mT (blue curve) applied field, to orient all magnetic domains and create a uniform magnetization. The signal originating from the magnetic circular dichroism indeed disappears close to 60~K. \\

For temperatures much lower than \Tc\ the transistor resistance becomes higher than the sensitivity of our instruments, preventing the evolution of the magnetic state to be probed. Out-of-plane transport measurements on \VIThree\ tunnel barriers do not suffer from this limitation~\cite{klein_probing_2018,song_giant_2018,wang_very_2018,kim_one_2018} and in the remaining of this manuscript, we focus on these measurements to investigate magnetism in \VIThree\ exfoliated multilayers. Representative $I-V$ characteristics of a \VIThree\ multilayer tunnel barrier are shown in Fig.~\ref{fig:3}a. They exhibit a very pronounced non-linearity consistent with Fowler-Nordheim tunneling  (see Fig.~\ref{fig:3}b): current flows when the applied bias tilts the bands in the \VIThree\ barriers and increases the tunneling transmission probability to a level that makes the tunneling current measurable, such that $\ln(I/V^2)$ depends linearly on $1/V$ at high bias~\cite{fowler_electron_1928}.\\

\begin{figure*}[ht]
    \centering
    \includegraphics[width=.95\textwidth]{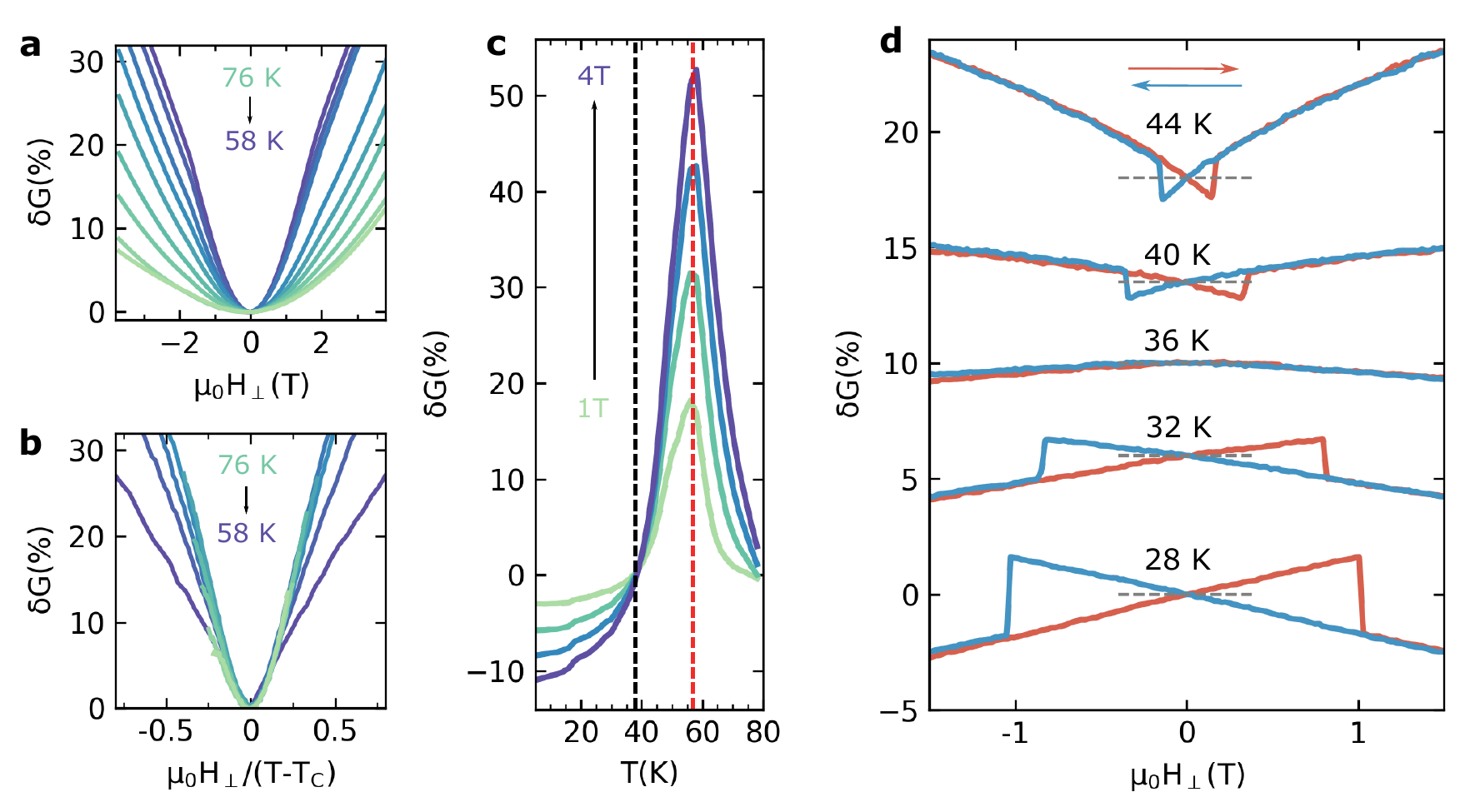}
    \caption{\textbf{(a)} Magnetoconductance curves $\delta G(\mu_0 H_\perp)$ for the temperature range between 76~K and 58~K, in 2~K steps. \textbf{(b)} same data as in panel  \textbf{(a)} plotted as a function of $\muHperp / (T-\Tc)$. At low magnetic fields, all the curves overlap, confirming the value of \Tc\ determined from transistor measurements. \textbf{(c)} Magnetoconductance as a function of temperature for different fixed magnetic fields \muHperp =1, 2, 3, and 4 T. The red dashed line at which  magnetoconductance is maximum corresponds to \Tc~=~57~K. The black dashed line marks the transition from positive to negative magnetoconductance ($T$~=~36~K). \textbf{(d)} Magnetoconductance traces as a function of magnetic field \muHperp\ for temperatures between $T$~=~44~K to $T$~=~28~K in 4~K steps. The traces are offset for clarity and the grey dashed line indicates the zero line for each trace, respectively. Red curves correspond to measurements performed with the magnetic field swept from negative to positive values;  blue curves correspond to data  taken by sweeping the field in the opposite direction.}    
    \label{fig:4}
\end{figure*}

Fig.~\ref{fig:3}c represents the tunneling magnetoconductance $\delta G (H_\perp,T) = (G(H_\perp,T)-G(0,T))/G(0,T)$ as a function of temperature, $T$, and magnetic field applied perpendicular to the layers, \muHperp, measured upon sweeping the field from either negative to positive (left) or from positive to negative (right) values. Because the $I-V$ curves are strongly non-linear, the absolute value of the conductance depends on applied bias, but the features observed in $\delta G(H_\perp, T)$ do not, \ie\ no qualitative aspect of the magnetoconductance on temperature and magnetic field dependence changes upon changing bias (see Supporting~Information~S3). The tunneling magnetoconductance can then be used to probe the properties of the magnetic state, as shown previously for \CrIThree, \CrClThree, \CrBrThree, and MnPS$_3$~ multilayers \cite{wang_magnetization_2021,wang_determining_2019,long_persistence_2020,wang_very_2018,kim_evolution_2019,song_giant_2018,klein_enhancement_2019,ghazaryan_magnon-assisted_2018}. Distinct features in $\delta G (H_\perp,T)$ can be identified near 60~K and 40~K. For $T$ close to 57~K, we observe a positive magnetoconductance exhibiting the same trends observed in our transistor devices, \ie\ the manifestation of the critical regime near a ferromagnetic phase transition that originates from the divergence of the magnetic susceptibility at \Tc. To confirm this conclusion, Fig.~\ref{fig:4}a,b  show magnetoconductance traces measured at different temperatures and the same data plotted as a function of $\mu_0 H_\perp /(T - \Tc)$ (using \Tc~=~57~K). All curves collapse again on top of each other for sufficiently small $H_\perp$, therefore, confirming that \Tc\ of thick exfoliated \VIThree\ multilayers is 57~K (or, again, between 56~K and 58~K, since magnetoconductance traces were measured with a 2~K interval), and not 50~K as in bulk crystals. \\

Well below \Tc, however, the tunneling magnetoconductance exhibits a behavior different from that seen in  \CrBrThree\ barriers. In \CrBrThree\ barriers, the tunneling magnetoconductance decreases and becomes vanishingly small at low temperature, but remains always positive~\cite{wang_magnetization_2021}. This is a direct consequence of the increase of the spontaneous magnetization of an Ising ferromagnet upon cooling: as the barrier is nearly completely spontaneously magnetized at low temperature, the application of a magnetic field increases the magnetization only slightly and has therefore little influence on the conductance. For \VIThree\ barriers, instead, when $T$ is lowered below \Tc\ the magnetoconductance vanishes at approximately $T$~=~36~K, and becomes negative upon further cooling, as shown in Fig.~\ref{fig:4}c. The change from positive to negative magnetoconductance is also apparent in Fig.~\ref{fig:4}d, which shows $\delta G$ as a function of $\mu_0 H_\perp$ for different values of $T$: indeed, the magnetoconductance increases upon increasing \muHperp\ for $T$~>~36~K and decreases upon increasing \muHperp\ for lower temperatures (the data also allow us to determine the coercive field, close to 3~T at $T$~=~4.2~K,  significantly larger than the value measured in bulk crystals, just above 1~T).\\

\begin{figure*}[ht!]
    \centering
    \includegraphics[width=0.9\textwidth]{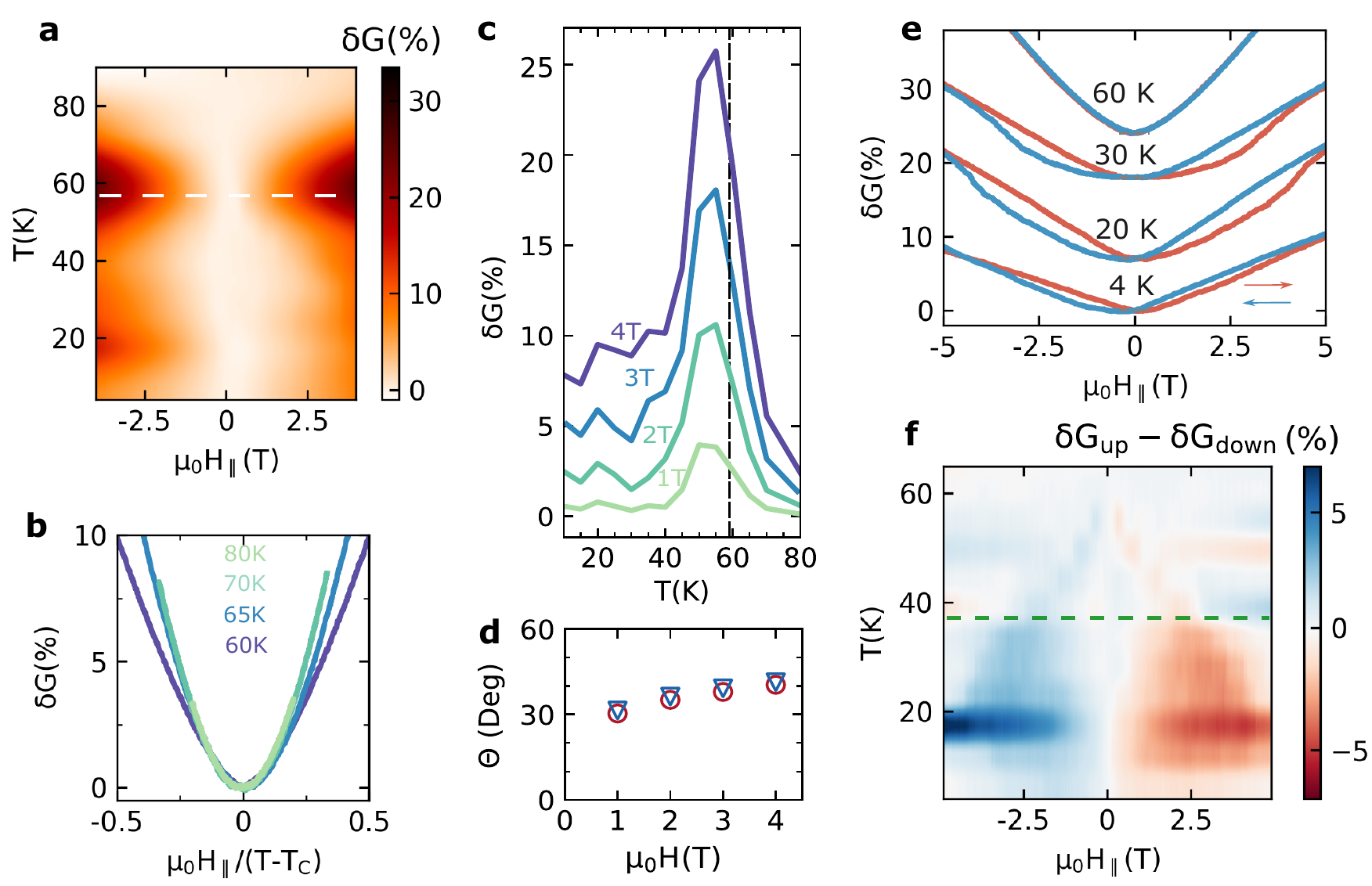}
    \caption{\textbf{(a)} Color plot of the tunneling magnetoconductance $\delta G$ as a function of temperature $T$ and magnetic field applied parallel \muHparallel\ to the \VIThree\ planes (data interpolated from magnetoconductance curves measured every 5~K). The two magnetoconductance \emph{lobes} centered around the Curie temperature (white dashed line indicates $T$~=~57~K) also develop in this case. \textbf{(b)} Magnetoconductance $\delta G$ as a function of \muHparallel$/(T-\Tc)$, measured at temperatures above the ferromagnetic transition (from 60~K to 80~K), showing the collapse of all curves at small applied field.  \textbf{(c)} Magnetoconductance as a function of temperature $\delta G(T)$ for fixed positive magnetic fields \muHparallel. The dashed line marks the Curie temperature \Tc~=~57~K. In contrast with the out-plane magnetoconductance, the in-plane magnetoconductance  remains always positive down to the lowest temperature investigated. \textbf{(d)} Calculated angle of the easy-axis $\Theta$ from measured values of $\delta G(H_{\parallel})$ and $\delta G(H_{\perp})$ at $T \sim$~\Tc\ as a function of the applied magnetic field $\mu_0 H$. Blue and red filled circles represent the angle for the 7~nm and 11~nm, respectively. \textbf{(e)} $\delta G(\mu_0 H_\parallel)$ curves for representative temperatures $T$, ranging from above to well below the Curie temperature (60~K, 30~K, 20~K and 4~K; the traces are offset for clarity). Solid red and blue curves represent the magnetoconductance measured while sweeping the magnetic field \muHparallel\ from negative to positive values or from positive to negative values, respectively. A small hysteresis is present at all temperature below \Tc, and it becomes very pronounced for temperatures close to 40~K, as shown in the color plot in \textbf{(f)}, which shows the difference between  magnetoconductance ($\delta G_{\rm{up}} - \delta G_{\rm{down}}$) measured while sweeping the field in opposite directions (the green dashed line corresponds to $T = 36$~K, \ie\ the temperature at which the magnetoconductance measured as a function of perpendicular field becomes negative, see Fig.~\ref{fig:2}; the plot shows an interpolation of magnetoconductance traces taken every 5~K).}
    \label{fig:5}
\end{figure*}
   
Finally, Fig~\ref{fig:5}a shows the dependence of the magnetoconductance $\delta G(\mu_0 H_\parallel, T)$ with magnetic field \muHparallel\ applied parallel to the \VIThree\ planes. Even in this case, we observe the characteristic manifestation of the critical regime near the ferromagnetic transition (see the \emph{lobes} of enhanced magnetoconductance for $T$ close to \Tc~=~57~K in Fig.~\ref{fig:5}a), and the low-field collapse of the magnetoconductance curves plotted versus $\mu_0 H_\parallel /(T - \Tc)$ (see Fig.~\ref{fig:5}b). Correspondingly,  $\delta G(\muHparallel, T)$ measured at fixed field as a function of $T$ peaks at \Tc, reaching values of 25~\%, comparable to, but smaller than, the values measured when the field is applied perpendicular to the layers. This difference carries important information because the relative intensity of the magnetoconductance peaks near \Tc\ measured with field applied perpendicular or parallel to the \VIThree\ layers allows determining the angle between the easy axis and the normal to the planes. \\

For a strongly anisotropic easy-axis ferromagnetic tunnel barrier just above \Tc\ (\ie\ in the paramagnetic state), the low-field magnetoconductance is proportional to the square of the magnetization, given by the magnetic susceptibility multiplied by the component of the applied field along the easy axis~\cite{wang_magnetization_2021}. For the same value of magnetic field applied either in-plane or perpendicular to the plane, the ratio $\delta G(H_{\parallel})/\delta G(H_{\perp})$  is thus given by the square of the ratio of the projections of the magnetic field along the easy axis, and can be used to calculate the angle $\Theta = \tan^{-1}(\sqrt{\delta G(H_{\parallel})/ \delta G({H_\perp})})$ between the easy axis and the direction normal to the \VIThree\ planes. The measurements should ideally be performed at small magnetic field (to ensure that the linear relation between $M$ and $H$ holds true), but with a field sufficiently large to have a signal well above the noise. In Figure~\ref{fig:5}d we show the derived angle for fields between 1 T and 4 T, for two different devices, giving  virtually identical results. We see that the angle between the easy axis and the direction normal to the VI$_3$ layers depends only weakly on field and approaches  30$\degree$ at 1 T, close to (but slightly smaller than) the one obtained from magnetization measurements performed on as-grown bulk crystals~\cite{koriki_magnetic_2021}.\\

Upon cooling below \Tc\ the magnetoconductance measured with field applied parallel to the plane decreases monotonously, and remains positive down to $T$~=~4~K. However, just under 40~K, a pronounced increase in hysteresis sets in (Fig.~\ref{fig:5}e-f), which was not reported for  \CrBrThree\ tunnel barriers~\cite{wang_magnetization_2021}. We conclude that for temperatures below 40~K  magnetotransport measurements  indicate a crossover to a regime  qualitatively different from that of an established Ising ferromagnet such as \CrBrThree~\cite{wang_magnetization_2021}. This crossover is signaled by a change in the sign of the magnetoconductance in perpendicular magnetic field and by a concomitant pronounced increase in the hysteresis of the magnetoconductance measured versus in-plane magnetic field. Such a crossover occurs in  the same temperature range in which an anomalous evolution of different magnetic properties takes place in bulk crystals,  signaling that the magnetic state of exfoliated layers and bulk crystals are closely related (despite the differences in structural properties, in critical temperature,  and coercive field). \\

The comparison of magnetotransport and magneto-optical experiments allows us to conclude that the observed phenomenology originates from the evolution of the in-plane component of the magnetization of \VIThree. Indeed, tunneling magnetotransport through ferromagnetic barriers probes the alignment of spins in the material, such that configurations of more aligned spins lead to larger tunneling conductance. The negative magnetoconductance observed in a perpendicular magnetic field, therefore, indicates that below 40~K the misalignment of the spins in the \VIThree\ multilayers increases rather than decreases upon increasing \muHperp. Since RMCD measurements  (the ones in Ref. \cite{lin_magnetism_2021} and ours, which are only sensitive to the $M_{\rm z}$ component of the magnetization~\cite{blundell_magnetism_2001}) find that  $M_{\rm z}$ keeps increasing when lowering $T$ below 40~K and upon applying a perpendicular magnetic field, the increased spin misalignment necessarily originates from the in-plane component of the magnetization.\\

The simplest scenario that we can envision is that the in-plane component of the magnetization in different \VIThree\ layers is not pointing in the same direction and that the relative angle between the in-plane component of the magnetization in adjacent layers depends on the applied perpendicular magnetic field and on temperature. In such a scenario, when $T$ is lowered below \Tc, the total spontaneous magnetization is small and is enhanced by the application of a magnetic field, leading to a positive magnetoconductance. However, as the temperature is lowered and the spontaneous magnetization builds up, this effect of the perpendicular magnetic field decreases, and for $T$~$<<$~\Tc\ its influence on the magnetoconductance becomes negligible (as found in \CrBrThree\ tunnel barriers~\cite{wang_magnetization_2021}). At that point, any misalignment of the in-plane component of the magnetization in different layers may start to dominate magnetotransport. Interestingly, this scenario provides a mechanism that may explain why the critical temperatures of mono and bilayer \VIThree\ (60~-~61~K) are slightly larger than that of thicker multilayers. Indeed, if the in-plane component of the magnetization in adjacent layers points in different directions, the effect of the intralayer exchange interaction responsible for the value of \Tc\ measured in isolated monolayers would be slightly weakened in multilayers, by the competition with the interlayer exchange interaction (the microscopic mechanism mediating the coupling between perpendicular magnetic field and in-plane component of the magnetization --implicit in the proposed scenario-- remains to be determined).\\

In summary, tunneling magnetoconductance measurements show that, despite having distinct crystallographic structures, exfoliated thin \VIThree\ crystals exhibit a magnetic state which resembles the one observed in bulk crystals, with the magnetic easy axis canted away from the normal to the VI$_3$ layers by a large angle. Even though our measurements do not allow the magnetic state to be determined, they show that an  Ising ferromagnetic state is not compatible with experimental observations and that deviations originate from the evolution of the in-plane component of the magnetization. These results illustrate the complementarity of magneto-optical and magnetotransport measurements to probe atomically thin 2D magnetic materials, since the commonly employed geometries of RMCD measurements are only sensitive to the component of the magnetization perpendicular to the layers, whereas tunneling magnetoconductance probe phenomena associated to the total magnetization. This conclusion is worth emphasizing because only a few techniques currently have sufficient sensitivity to probe the magnetic properties of atomically thin layers, and it is important to understand which magnetic properties are probed by different measurements.

\section*{Acknowledgements}
The authors gratefully acknowledge Alexandre Ferreira for technical support, Qian Song for help with sample preparation, and Ignacio Gutiérrez-Lezama and Marco Gibertini for fruitful discussions. A. F. Morpurgo gratefully acknowledges the Swiss National  Science Foundation and the EU Graphene Flagship project for support.   J. Li, C.A. Occhialini and  R. Comin  acknowledge support by the STC Center for Integrated Quantum Materials, NSF Grant No. DMR-1231319.\\

\providecommand{\latin}[1]{#1}
\makeatletter
\providecommand{\doi}
  {\begingroup\let\do\@makeother\dospecials
  \catcode`\{=1 \catcode`\}=2 \doi@aux}
\providecommand{\doi@aux}[1]{\endgroup\texttt{#1}}
\makeatother
\providecommand*\mcitethebibliography{\thebibliography}
\csname @ifundefined\endcsname{endmcitethebibliography}
  {\let\endmcitethebibliography\endthebibliography}{}


\begin{mcitethebibliography}{39}
\providecommand*\natexlab[1]{#1}
\providecommand*\mciteSetBstSublistMode[1]{}
\providecommand*\mciteSetBstMaxWidthForm[2]{}
\providecommand*\mciteBstWouldAddEndPuncttrue
  {\def\EndOfBibitem{\unskip.}}
\providecommand*\mciteBstWouldAddEndPunctfalse
  {\let\EndOfBibitem\relax}
\providecommand*\mciteSetBstMidEndSepPunct[3]{}
\providecommand*\mciteSetBstSublistLabelBeginEnd[3]{}
\providecommand*\EndOfBibitem{}
\mciteSetBstSublistMode{f}
\mciteSetBstMaxWidthForm{subitem}{(\alph{mcitesubitemcount})}
\mciteSetBstSublistLabelBeginEnd
  {\mcitemaxwidthsubitemform\space}
  {\relax}
  {\relax}

\bibitem[Novoselov \latin{et~al.}(2005)Novoselov, Jiang, Schedin, Booth,
  Khotkevich, Morozov, and Geim]{novoselov_two-dimensional_2005}
Novoselov,~K.~S.; Jiang,~D.; Schedin,~F.; Booth,~T.~J.; Khotkevich,~V.~V.;
  Morozov,~S.~V.; Geim,~A.~K. Two-dimensional atomic crystals.
  \emph{Proceedings of the National Academy of Sciences} \textbf{2005},
  \emph{102}, 10451--10453, Publisher: Proceedings of the National Academy of
  Sciences\relax
\mciteBstWouldAddEndPuncttrue
\mciteSetBstMidEndSepPunct{\mcitedefaultmidpunct}
{\mcitedefaultendpunct}{\mcitedefaultseppunct}\relax
\EndOfBibitem
\bibitem[Mounet \latin{et~al.}(2018)Mounet, Gibertini, Schwaller, Campi,
  Merkys, Marrazzo, Sohier, Castelli, Cepellotti, Pizzi, and
  Marzari]{mounet_two-dimensional_2018}
Mounet,~N.; Gibertini,~M.; Schwaller,~P.; Campi,~D.; Merkys,~A.; Marrazzo,~A.;
  Sohier,~T.; Castelli,~I.~E.; Cepellotti,~A.; Pizzi,~G.; Marzari,~N.
  Two-dimensional materials from high-throughput computational exfoliation of
  experimentally known compounds. \emph{Nature Nanotechnology} \textbf{2018},
  \emph{13}, 246--252, Number: 3 Publisher: Nature Publishing Group\relax
\mciteBstWouldAddEndPuncttrue
\mciteSetBstMidEndSepPunct{\mcitedefaultmidpunct}
{\mcitedefaultendpunct}{\mcitedefaultseppunct}\relax
\EndOfBibitem
\bibitem[Gong \latin{et~al.}(2017)Gong, Li, Li, Ji, Stern, Xia, Cao, Bao, Wang,
  Wang, Qiu, Cava, Louie, Xia, and Zhang]{gong_discovery_2017}
Gong,~C.; Li,~L.; Li,~Z.; Ji,~H.; Stern,~A.; Xia,~Y.; Cao,~T.; Bao,~W.;
  Wang,~C.; Wang,~Y.; Qiu,~Z.~Q.; Cava,~R.~J.; Louie,~S.~G.; Xia,~J.; Zhang,~X.
  Discovery of intrinsic ferromagnetism in two-dimensional van der {Waals}
  crystals. \emph{Nature} \textbf{2017}, \emph{546}, 265--269\relax
\mciteBstWouldAddEndPuncttrue
\mciteSetBstMidEndSepPunct{\mcitedefaultmidpunct}
{\mcitedefaultendpunct}{\mcitedefaultseppunct}\relax
\EndOfBibitem
\bibitem[Huang \latin{et~al.}(2017)Huang, Clark, Navarro-Moratalla, Klein,
  Cheng, Seyler, Zhong, Schmidgall, McGuire, Cobden, Yao, Xiao,
  Jarillo-Herrero, and Xu]{huang_layer-dependent_2017}
Huang,~B.; Clark,~G.; Navarro-Moratalla,~E.; Klein,~D.~R.; Cheng,~R.;
  Seyler,~K.~L.; Zhong,~D.; Schmidgall,~E.; McGuire,~M.~A.; Cobden,~D.~H.;
  Yao,~W.; Xiao,~D.; Jarillo-Herrero,~P.; Xu,~X. Layer-dependent ferromagnetism
  in a van der {{Waals}} crystal down to the monolayer limit. \emph{Nature}
  \textbf{2017}, \emph{546}, 270--273\relax
\mciteBstWouldAddEndPuncttrue
\mciteSetBstMidEndSepPunct{\mcitedefaultmidpunct}
{\mcitedefaultendpunct}{\mcitedefaultseppunct}\relax
\EndOfBibitem
\bibitem[Burch \latin{et~al.}(2018)Burch, Mandrus, and
  Park]{burch_magnetism_2018}
Burch,~K.~S.; Mandrus,~D.; Park,~J.-G. Magnetism in two-dimensional van der
  {Waals} materials. \emph{Nature} \textbf{2018}, \emph{563}, 47--52\relax
\mciteBstWouldAddEndPuncttrue
\mciteSetBstMidEndSepPunct{\mcitedefaultmidpunct}
{\mcitedefaultendpunct}{\mcitedefaultseppunct}\relax
\EndOfBibitem
\bibitem[Gong and Zhang(2019)Gong, and Zhang]{gong_two-dimensional_2019}
Gong,~C.; Zhang,~X. Two-dimensional magnetic crystals and emergent
  heterostructure devices. \emph{Science} \textbf{2019}, \emph{363},
  eaav4450\relax
\mciteBstWouldAddEndPuncttrue
\mciteSetBstMidEndSepPunct{\mcitedefaultmidpunct}
{\mcitedefaultendpunct}{\mcitedefaultseppunct}\relax
\EndOfBibitem
\bibitem[Gibertini \latin{et~al.}(2019)Gibertini, Koperski, Morpurgo, and
  Novoselov]{gibertini_magnetic_2019}
Gibertini,~M.; Koperski,~M.; Morpurgo,~A.~F.; Novoselov,~K.~S. Magnetic {2D}
  materials and heterostructures. \emph{Nature Nanotechnology} \textbf{2019},
  \emph{14}, 408--419\relax
\mciteBstWouldAddEndPuncttrue
\mciteSetBstMidEndSepPunct{\mcitedefaultmidpunct}
{\mcitedefaultendpunct}{\mcitedefaultseppunct}\relax
\EndOfBibitem
\bibitem[Mak \latin{et~al.}(2019)Mak, Shan, and Ralph]{mak_probing_2019}
Mak,~K.~F.; Shan,~J.; Ralph,~D.~C. Probing and controlling magnetic states in
  {2D} layered magnetic materials. \emph{Nature Reviews Physics} \textbf{2019},
  \emph{1}, 646--661\relax
\mciteBstWouldAddEndPuncttrue
\mciteSetBstMidEndSepPunct{\mcitedefaultmidpunct}
{\mcitedefaultendpunct}{\mcitedefaultseppunct}\relax
\EndOfBibitem
\bibitem[Huang \latin{et~al.}(2020)Huang, McGuire, May, Xiao, Jarillo-Herrero,
  and Xu]{huang_emergent_2020}
Huang,~B.; McGuire,~M.~A.; May,~A.~F.; Xiao,~D.; Jarillo-Herrero,~P.; Xu,~X.
  Emergent phenomena and proximity effects in two-dimensional magnets and
  heterostructures. \emph{Nature Materials} \textbf{2020}, \emph{19},
  1276--1289\relax
\mciteBstWouldAddEndPuncttrue
\mciteSetBstMidEndSepPunct{\mcitedefaultmidpunct}
{\mcitedefaultendpunct}{\mcitedefaultseppunct}\relax
\EndOfBibitem
\bibitem[Sun \latin{et~al.}(2019)Sun, Yi, Song, Clark, Huang, Shan, Wu, Huang,
  Gao, Chen, McGuire, Cao, Xiao, Liu, Yao, Xu, and Wu]{sun_giant_2019}
Sun,~Z. \latin{et~al.}  Giant nonreciprocal second-harmonic generation from
  antiferromagnetic bilayer {CrI3}. \emph{Nature} \textbf{2019}, \emph{572},
  497--501\relax
\mciteBstWouldAddEndPuncttrue
\mciteSetBstMidEndSepPunct{\mcitedefaultmidpunct}
{\mcitedefaultendpunct}{\mcitedefaultseppunct}\relax
\EndOfBibitem
\bibitem[Klein \latin{et~al.}(2019)Klein, MacNeill, Song, Larson, Fang, Xu,
  Ribeiro, Canfield, Kaxiras, Comin, and
  Jarillo-Herrero]{klein_enhancement_2019}
Klein,~D.~R.; MacNeill,~D.; Song,~Q.; Larson,~D.~T.; Fang,~S.; Xu,~M.;
  Ribeiro,~R.~A.; Canfield,~P.~C.; Kaxiras,~E.; Comin,~R.; Jarillo-Herrero,~P.
  Enhancement of interlayer exchange in an ultrathin two-dimensional magnet.
  \emph{Nature Physics} \textbf{2019}, \emph{15}, 1255--1260\relax
\mciteBstWouldAddEndPuncttrue
\mciteSetBstMidEndSepPunct{\mcitedefaultmidpunct}
{\mcitedefaultendpunct}{\mcitedefaultseppunct}\relax
\EndOfBibitem
\bibitem[Ubrig \latin{et~al.}(2019)Ubrig, Wang, Teyssier, Taniguchi, Watanabe,
  Giannini, Morpurgo, and Gibertini]{ubrig_low-temperature_2019}
Ubrig,~N.; Wang,~Z.; Teyssier,~J.; Taniguchi,~T.; Watanabe,~K.; Giannini,~E.;
  Morpurgo,~A.~F.; Gibertini,~M. Low-temperature monoclinic layer stacking in
  atomically thin {CrI}$_{\textrm{3}}$\$ crystals. \emph{2D Materials}
  \textbf{2019}, \emph{7}, 015007\relax
\mciteBstWouldAddEndPuncttrue
\mciteSetBstMidEndSepPunct{\mcitedefaultmidpunct}
{\mcitedefaultendpunct}{\mcitedefaultseppunct}\relax
\EndOfBibitem
\bibitem[McCreary \latin{et~al.}(2020)McCreary, Mai, Utermohlen, Simpson,
  Garrity, Feng, Shcherbakov, Zhu, Hu, Weber, Watanabe, Taniguchi, Goldberger,
  Mao, Lau, Lu, Trivedi, Valdés~Aguilar, and
  Hight~Walker]{mccreary_distinct_2020}
McCreary,~A. \latin{et~al.}  Distinct magneto-{Raman} signatures of spin-flip
  phase transitions in {CrI3}. \emph{Nature Communications} \textbf{2020},
  \emph{11}, 3879, Number: 1 Publisher: Nature Publishing Group\relax
\mciteBstWouldAddEndPuncttrue
\mciteSetBstMidEndSepPunct{\mcitedefaultmidpunct}
{\mcitedefaultendpunct}{\mcitedefaultseppunct}\relax
\EndOfBibitem
\bibitem[Gibertini(2020)]{gibertini_magnetism_2020}
Gibertini,~M. Magnetism and stability of all primitive stacking patterns in
  bilayer chromium trihalides. \emph{Journal of Physics D: Applied Physics}
  \textbf{2020}, \emph{54}, 064002, Publisher: IOP Publishing\relax
\mciteBstWouldAddEndPuncttrue
\mciteSetBstMidEndSepPunct{\mcitedefaultmidpunct}
{\mcitedefaultendpunct}{\mcitedefaultseppunct}\relax
\EndOfBibitem
\bibitem[Dillon \latin{et~al.}(1966)Dillon, Kamimura, and
  Remeika]{dillon_magneto-optical_1966}
Dillon,~J.~F.; Kamimura,~H.; Remeika,~J.~P. Magneto-optical properties of
  ferromagnetic chromium trihalides. \emph{Journal of Physics and Chemistry of
  Solids} \textbf{1966}, \emph{27}, 1531--1549\relax
\mciteBstWouldAddEndPuncttrue
\mciteSetBstMidEndSepPunct{\mcitedefaultmidpunct}
{\mcitedefaultendpunct}{\mcitedefaultseppunct}\relax
\EndOfBibitem
\bibitem[Wang \latin{et~al.}(2018)Wang, Gutiérrez-Lezama, Ubrig, Kroner,
  Gibertini, Taniguchi, Watanabe, Imamoğlu, Giannini, and
  Morpurgo]{wang_very_2018}
Wang,~Z.; Gutiérrez-Lezama,~I.; Ubrig,~N.; Kroner,~M.; Gibertini,~M.;
  Taniguchi,~T.; Watanabe,~K.; Imamoğlu,~A.; Giannini,~E.; Morpurgo,~A.~F.
  Very large tunneling magnetoresistance in layered magnetic semiconductor
  {CrI3}. \emph{Nature Communications} \textbf{2018}, \emph{9}, 2516\relax
\mciteBstWouldAddEndPuncttrue
\mciteSetBstMidEndSepPunct{\mcitedefaultmidpunct}
{\mcitedefaultendpunct}{\mcitedefaultseppunct}\relax
\EndOfBibitem
\bibitem[Wang \latin{et~al.}(2019)Wang, Gibertini, Dumcenco, Taniguchi,
  Watanabe, Giannini, and Morpurgo]{wang_determining_2019}
Wang,~Z.; Gibertini,~M.; Dumcenco,~D.; Taniguchi,~T.; Watanabe,~K.;
  Giannini,~E.; Morpurgo,~A.~F. Determining the phase diagram of atomically
  thin layered antiferromagnet {CrCl3}. \emph{Nature Nanotechnology}
  \textbf{2019}, \emph{14}, 1116--1122\relax
\mciteBstWouldAddEndPuncttrue
\mciteSetBstMidEndSepPunct{\mcitedefaultmidpunct}
{\mcitedefaultendpunct}{\mcitedefaultseppunct}\relax
\EndOfBibitem
\bibitem[Son \latin{et~al.}(2019)Son, Coak, Lee, Kim, Kim, Hamidov, Cho, Liu,
  Jarvis, Brown, Kim, Park, Khomskii, Saxena, and Park]{son_bulk_2019}
Son,~S.; Coak,~M.~J.; Lee,~N.; Kim,~J.; Kim,~T.~Y.; Hamidov,~H.; Cho,~H.;
  Liu,~C.; Jarvis,~D.~M.; Brown,~P. A.~C.; Kim,~J.~H.; Park,~C.-H.;
  Khomskii,~D.~I.; Saxena,~S.~S.; Park,~J.-G. Bulk properties of the van der
  {Waals} hard ferromagnet {VI} 3. \emph{Physical Review B} \textbf{2019},
  \emph{99}, 041402\relax
\mciteBstWouldAddEndPuncttrue
\mciteSetBstMidEndSepPunct{\mcitedefaultmidpunct}
{\mcitedefaultendpunct}{\mcitedefaultseppunct}\relax
\EndOfBibitem
\bibitem[Lin \latin{et~al.}(2021)Lin, Huang, Hwangbo, Jiang, Zhang, Liu, Fei,
  Lv, Millis, McGuire, Xiao, Chu, and Xu]{lin_magnetism_2021}
Lin,~Z.; Huang,~B.; Hwangbo,~K.; Jiang,~Q.; Zhang,~Q.; Liu,~Z.; Fei,~Z.;
  Lv,~H.; Millis,~A.; McGuire,~M.; Xiao,~D.; Chu,~J.-H.; Xu,~X. Magnetism and
  {Its} {Structural} {Coupling} {Effects} in {2D} {Ising} {Ferromagnetic}
  {Insulator} {VI}3. \emph{Nano Letters} \textbf{2021}, \emph{21},
  9180--9186\relax
\mciteBstWouldAddEndPuncttrue
\mciteSetBstMidEndSepPunct{\mcitedefaultmidpunct}
{\mcitedefaultendpunct}{\mcitedefaultseppunct}\relax
\EndOfBibitem
\bibitem[Doležal \latin{et~al.}(2019)Doležal, Kratochvílová, Holý,
  Čermák, Sechovský, Dušek, Míšek, Chakraborty, Noda, Son, and
  Park]{dolezal_crystal_2019}
Doležal,~P.; Kratochvílová,~M.; Holý,~V.; Čermák,~P.; Sechovský,~V.;
  Dušek,~M.; Míšek,~M.; Chakraborty,~T.; Noda,~Y.; Son,~S.; Park,~J.-G.
  Crystal structures and phase transitions of the van der {Waals} ferromagnet
  {V} {I} 3. \emph{Physical Review Materials} \textbf{2019}, \emph{3},
  121401\relax
\mciteBstWouldAddEndPuncttrue
\mciteSetBstMidEndSepPunct{\mcitedefaultmidpunct}
{\mcitedefaultendpunct}{\mcitedefaultseppunct}\relax
\EndOfBibitem
\bibitem[Koriki \latin{et~al.}(2021)Koriki, Míšek, Pospíšil,
  Kratochvílová, Carva, Prokleška, Doležal, Kaštil, Son, Park, and
  Sechovský]{koriki_magnetic_2021}
Koriki,~A.; Míšek,~M.; Pospíšil,~J.; Kratochvílová,~M.; Carva,~K.;
  Prokleška,~J.; Doležal,~P.; Kaštil,~J.; Son,~S.; Park,~J.-G.;
  Sechovský,~V. Magnetic anisotropy in the van der {Waals} ferromagnet {V} {I}
  3. \emph{Physical Review B} \textbf{2021}, \emph{103}, 174401\relax
\mciteBstWouldAddEndPuncttrue
\mciteSetBstMidEndSepPunct{\mcitedefaultmidpunct}
{\mcitedefaultendpunct}{\mcitedefaultseppunct}\relax
\EndOfBibitem
\bibitem[Gati \latin{et~al.}(2019)Gati, Inagaki, Kong, Cava, Furukawa,
  Canfield, and Bud'ko]{gati_multiple_2019}
Gati,~E.; Inagaki,~Y.; Kong,~T.; Cava,~R.~J.; Furukawa,~Y.; Canfield,~P.~C.;
  Bud'ko,~S.~L. Multiple ferromagnetic transitions and structural distortion in
  the van der {Waals} ferromagnet {VI} 3 at ambient and finite pressures.
  \emph{Physical Review B} \textbf{2019}, \emph{100}, 094408\relax
\mciteBstWouldAddEndPuncttrue
\mciteSetBstMidEndSepPunct{\mcitedefaultmidpunct}
{\mcitedefaultendpunct}{\mcitedefaultseppunct}\relax
\EndOfBibitem
\bibitem[Kong \latin{et~al.}(2019)Kong, Stolze, Timmons, Tao, Ni, Guo, Yang,
  Prozorov, and Cava]{kong_vi_2019}
Kong,~T.; Stolze,~K.; Timmons,~E.~I.; Tao,~J.; Ni,~D.; Guo,~S.; Yang,~Z.;
  Prozorov,~R.; Cava,~R.~J. {VI}3 —a {New} {Layered} {Ferromagnetic}
  {Semiconductor}. \emph{Advanced Materials} \textbf{2019}, \emph{31},
  1808074\relax
\mciteBstWouldAddEndPuncttrue
\mciteSetBstMidEndSepPunct{\mcitedefaultmidpunct}
{\mcitedefaultendpunct}{\mcitedefaultseppunct}\relax
\EndOfBibitem
\bibitem[Tian \latin{et~al.}(2019)Tian, Zhang, Li, Ying, Li, Zhang, Liu, and
  Lei]{tian_ferromagnetic_2019}
Tian,~S.; Zhang,~J.-F.; Li,~C.; Ying,~T.; Li,~S.; Zhang,~X.; Liu,~K.; Lei,~H.
  Ferromagnetic van der {Waals} {Crystal} {VI}3. \emph{Journal of the American
  Chemical Society} \textbf{2019}, \emph{141}, 5326--5333\relax
\mciteBstWouldAddEndPuncttrue
\mciteSetBstMidEndSepPunct{\mcitedefaultmidpunct}
{\mcitedefaultendpunct}{\mcitedefaultseppunct}\relax
\EndOfBibitem
\bibitem[Valenta \latin{et~al.}(2021)Valenta, Kratochvílová, Míšek, Carva,
  Kaštil, Doležal, Opletal, Čermák, Proschek, Uhlířová, Prchal, Coak,
  Son, Park, and Sechovský]{valenta_pressure-induced_2021}
Valenta,~J.; Kratochvílová,~M.; Míšek,~M.; Carva,~K.; Kaštil,~J.;
  Doležal,~P.; Opletal,~P.; Čermák,~P.; Proschek,~P.; Uhlířová,~K.;
  Prchal,~J.; Coak,~M.~J.; Son,~S.; Park,~J.-G.; Sechovský,~V.
  Pressure-induced large increase of {Curie} temperature of the van der {Waals}
  ferromagnet {V} {I} 3. \emph{Physical Review B} \textbf{2021}, \emph{103},
  054424\relax
\mciteBstWouldAddEndPuncttrue
\mciteSetBstMidEndSepPunct{\mcitedefaultmidpunct}
{\mcitedefaultendpunct}{\mcitedefaultseppunct}\relax
\EndOfBibitem
\bibitem[Marchandier \latin{et~al.}(2021)Marchandier, Dubouis, Fauth, Avdeev,
  Grimaud, Tarascon, and Rousse]{marchandier_crystallographic_2021}
Marchandier,~T.; Dubouis,~N.; Fauth,~F.; Avdeev,~M.; Grimaud,~A.;
  Tarascon,~J.-M.; Rousse,~G. Crystallographic and magnetic structures of the
  {VI} 3 and {LiVI} 3 van der {Waals} compounds. \emph{Physical Review B}
  \textbf{2021}, \emph{104}, 014105\relax
\mciteBstWouldAddEndPuncttrue
\mciteSetBstMidEndSepPunct{\mcitedefaultmidpunct}
{\mcitedefaultendpunct}{\mcitedefaultseppunct}\relax
\EndOfBibitem
\bibitem[Lyu \latin{et~al.}(2020)Lyu, Gao, Zhang, Wang, Wu, Chen, Zhang, Li,
  Huang, Zhang, Chen, Mei, Yan, Zhao, Huang, and Huang]{lyu_probing_2020}
Lyu,~B. \latin{et~al.}  Probing the {Ferromagnetism} and {Spin} {Wave} {Gap} in
  {VI}3 by {Helicity}-{Resolved} {Raman} {Spectroscopy}. \emph{Nano Letters}
  \textbf{2020}, \emph{20}, 6024--6031\relax
\mciteBstWouldAddEndPuncttrue
\mciteSetBstMidEndSepPunct{\mcitedefaultmidpunct}
{\mcitedefaultendpunct}{\mcitedefaultseppunct}\relax
\EndOfBibitem
\bibitem[Broadway \latin{et~al.}(2020)Broadway, Scholten, Tan, Dontschuk,
  Lillie, Johnson, Zheng, Wang, Oganov, Tian, Li, Lei, Wang, Hollenberg, and
  Tetienne]{broadway_imaging_2020}
Broadway,~D.~A.; Scholten,~S.~C.; Tan,~C.; Dontschuk,~N.; Lillie,~S.~E.;
  Johnson,~B.~C.; Zheng,~G.; Wang,~Z.; Oganov,~A.~R.; Tian,~S.; Li,~C.;
  Lei,~H.; Wang,~L.; Hollenberg,~L. C.~L.; Tetienne,~J. Imaging {Domain}
  {Reversal} in an {Ultrathin} {Van} der {Waals} {Ferromagnet}. \emph{Advanced
  Materials} \textbf{2020}, \emph{32}, 2003314\relax
\mciteBstWouldAddEndPuncttrue
\mciteSetBstMidEndSepPunct{\mcitedefaultmidpunct}
{\mcitedefaultendpunct}{\mcitedefaultseppunct}\relax
\EndOfBibitem
\bibitem[Kim \latin{et~al.}(2019)Kim, Yang, Li, Jiang, Jin, Tao, Nichols,
  Sfigakis, Zhong, Li, Tian, Cory, Miao, Shan, Mak, Lei, Sun, Zhao, and
  Tsen]{kim_evolution_2019}
Kim,~H.~H. \latin{et~al.}  Evolution of interlayer and intralayer magnetism in
  three atomically thin chromium trihalides. \emph{Proceedings of the National
  Academy of Sciences} \textbf{2019}, \emph{116}, 11131--11136\relax
\mciteBstWouldAddEndPuncttrue
\mciteSetBstMidEndSepPunct{\mcitedefaultmidpunct}
{\mcitedefaultendpunct}{\mcitedefaultseppunct}\relax
\EndOfBibitem
\bibitem[Wang \latin{et~al.}(2021)Wang, Gutiérrez-Lezama, Dumcenco, Ubrig,
  Taniguchi, Watanabe, Giannini, Gibertini, and
  Morpurgo]{wang_magnetization_2021}
Wang,~Z.; Gutiérrez-Lezama,~I.; Dumcenco,~D.; Ubrig,~N.; Taniguchi,~T.;
  Watanabe,~K.; Giannini,~E.; Gibertini,~M.; Morpurgo,~A.~F. Magnetization
  dependent tunneling conductance of ferromagnetic barriers. \emph{Nature
  Communications} \textbf{2021}, \emph{12}, 6659\relax
\mciteBstWouldAddEndPuncttrue
\mciteSetBstMidEndSepPunct{\mcitedefaultmidpunct}
{\mcitedefaultendpunct}{\mcitedefaultseppunct}\relax
\EndOfBibitem
\bibitem[Zomer \latin{et~al.}(2011)Zomer, Dash, Tombros, and van
  Wees]{zomer_transfer_2011}
Zomer,~P.~J.; Dash,~S.~P.; Tombros,~N.; van Wees,~B.~J. A transfer technique
  for high mobility graphene devices on commercially available hexagonal boron
  nitride. \emph{Appl. Phys. Lett.} \textbf{2011}, \emph{99}, 232104\relax
\mciteBstWouldAddEndPuncttrue
\mciteSetBstMidEndSepPunct{\mcitedefaultmidpunct}
{\mcitedefaultendpunct}{\mcitedefaultseppunct}\relax
\EndOfBibitem
\bibitem[Klein \latin{et~al.}(2018)Klein, MacNeill, Lado, Soriano,
  Navarro-Moratalla, Watanabe, Taniguchi, Manni, Canfield, Fernández-Rossier,
  and Jarillo-Herrero]{klein_probing_2018}
Klein,~D.~R.; MacNeill,~D.; Lado,~J.~L.; Soriano,~D.; Navarro-Moratalla,~E.;
  Watanabe,~K.; Taniguchi,~T.; Manni,~S.; Canfield,~P.; Fernández-Rossier,~J.;
  Jarillo-Herrero,~P. Probing magnetism in {2D} van der {Waals} crystalline
  insulators via electron tunneling. \emph{Science} \textbf{2018},
  eaar3617\relax
\mciteBstWouldAddEndPuncttrue
\mciteSetBstMidEndSepPunct{\mcitedefaultmidpunct}
{\mcitedefaultendpunct}{\mcitedefaultseppunct}\relax
\EndOfBibitem
\bibitem[Song \latin{et~al.}(2018)Song, Cai, Tu, Zhang, Huang, Wilson, Seyler,
  Zhu, Taniguchi, Watanabe, McGuire, Cobden, Xiao, Yao, and
  Xu]{song_giant_2018}
Song,~T.; Cai,~X.; Tu,~M. W.-Y.; Zhang,~X.; Huang,~B.; Wilson,~N.~P.;
  Seyler,~K.~L.; Zhu,~L.; Taniguchi,~T.; Watanabe,~K.; McGuire,~M.~A.;
  Cobden,~D.~H.; Xiao,~D.; Yao,~W.; Xu,~X. Giant tunneling magnetoresistance in
  spin-filter van der {Waals} heterostructures. \emph{Science} \textbf{2018},
  eaar4851\relax
\mciteBstWouldAddEndPuncttrue
\mciteSetBstMidEndSepPunct{\mcitedefaultmidpunct}
{\mcitedefaultendpunct}{\mcitedefaultseppunct}\relax
\EndOfBibitem
\bibitem[Kim \latin{et~al.}(2018)Kim, Yang, Patel, Sfigakis, Li, Tian, Lei, and
  Tsen]{kim_one_2018}
Kim,~H.~H.; Yang,~B.; Patel,~T.; Sfigakis,~F.; Li,~C.; Tian,~S.; Lei,~H.;
  Tsen,~A.~W. One {Million} {Percent} {Tunnel} {Magnetoresistance} in a
  {Magnetic} van der {Waals} {Heterostructure}. \emph{Nano Letters}
  \textbf{2018}, \emph{18}, 4885--4890\relax
\mciteBstWouldAddEndPuncttrue
\mciteSetBstMidEndSepPunct{\mcitedefaultmidpunct}
{\mcitedefaultendpunct}{\mcitedefaultseppunct}\relax
\EndOfBibitem
\bibitem[Fowler and Nordheim(1928)Fowler, and Nordheim]{fowler_electron_1928}
Fowler,~R.~H.; Nordheim,~L. Electron emission in intense electric fields.
  \emph{Proceedings of the Royal Society of London. Series A, Containing Papers
  of a Mathematical and Physical Character} \textbf{1928}, \emph{119},
  173--181\relax
\mciteBstWouldAddEndPuncttrue
\mciteSetBstMidEndSepPunct{\mcitedefaultmidpunct}
{\mcitedefaultendpunct}{\mcitedefaultseppunct}\relax
\EndOfBibitem
\bibitem[Long \latin{et~al.}(2020)Long, Henck, Gibertini, Dumcenco, Wang,
  Taniguchi, Watanabe, Giannini, and Morpurgo]{long_persistence_2020}
Long,~G.; Henck,~H.; Gibertini,~M.; Dumcenco,~D.; Wang,~Z.; Taniguchi,~T.;
  Watanabe,~K.; Giannini,~E.; Morpurgo,~A.~F. Persistence of {Magnetism} in
  {Atomically} {Thin} {MnPS3} {Crystals}. \emph{Nano Letters} \textbf{2020},
  \emph{20}, 2452--2459\relax
\mciteBstWouldAddEndPuncttrue
\mciteSetBstMidEndSepPunct{\mcitedefaultmidpunct}
{\mcitedefaultendpunct}{\mcitedefaultseppunct}\relax
\EndOfBibitem
\bibitem[Ghazaryan \latin{et~al.}(2018)Ghazaryan, Greenaway, Wang,
  Guarochico-Moreira, Vera-Marun, Yin, Liao, Morozov, Kristanovski,
  Lichtenstein, Katsnelson, Withers, Mishchenko, Eaves, Geim, Novoselov, and
  Misra]{ghazaryan_magnon-assisted_2018}
Ghazaryan,~D. \latin{et~al.}  Magnon-assisted tunnelling in van der {Waals}
  heterostructures based on {CrBr3}. \emph{Nature Electronics} \textbf{2018},
  \emph{1}, 344--349\relax
\mciteBstWouldAddEndPuncttrue
\mciteSetBstMidEndSepPunct{\mcitedefaultmidpunct}
{\mcitedefaultendpunct}{\mcitedefaultseppunct}\relax
\EndOfBibitem
\bibitem[Blundell(2001)]{blundell_magnetism_2001}
Blundell,~S. \emph{Magnetism in {Condensed} {Matter}}; OUP Oxford, 2001\relax
\mciteBstWouldAddEndPuncttrue
\mciteSetBstMidEndSepPunct{\mcitedefaultmidpunct}
{\mcitedefaultendpunct}{\mcitedefaultseppunct}\relax
\EndOfBibitem
\end{mcitethebibliography}
\end{document}